\documentclass[twocolumn,twocolappendix]{aastex701}

\usepackage{amsmath}

\newcommand{\heilabel}{\ion{He}{1}~$\lambda$10830}
\newcommand{\hei}{\ion{He}{1}}
\newcommand{\pag}{Pa$\gamma$}
\newcommand{\hii}{\ion{H}{2}}
\newcommand{\kms}{km~s$^{-1}$}
\newcommand{\term}{2\,$^3{\rm S}_1$}
\newcommand{\vlsrpag}{$v_{\rm LSR}$(\pag)}

\defcitealias{Draine2026ApJ}{D26}
\defcitealias{Pellegrini2007ApJ}{P07}
\defcitealias{astropy2022}{Astropy Collaboration et al. 2022}

\graphicspath{{./}{figures/}}

\begin{document}

\title{He I $\lambda$10830 Resonant Scattering in M17: Toward a Diagnostic of Optical Depth and Gas Kinematics}

\author[orcid=0000-0001-6401-723X,gname='Hiroaki',sname='Sameshima']{Hiroaki Sameshima}
\affiliation{Institute of Astronomy, Graduate School of Science, The University of Tokyo, 2-21-1 Osawa, Mitaka, Tokyo 181-0015, Japan}
\email[show]{hsameshima@ioa.s.u-tokyo.ac.jp}

\author[orcid=0000-0002-2861-4069]{Daisuke Taniguchi}
\altaffiliation{The Tokyo Center For Excellence Project, Tokyo Metropolitan University}
\affiliation{Department of Physics, Tokyo Metropolitan University, 1-1 Minami-Osawa, Hachioji, Tokyo 192-0397, Japan}
\affiliation{Laboratory of Infrared High-resolution spectroscopy (LiH), Koyama Astronomical Observatory, Kyoto Sangyo University, Motoyama, Kamigamo, Kita-ku, Kyoto 603-8555, Japan}
\email{d.taniguchi.astro@gmail.com}

\author[orcid=0009-0003-6827-6396]{Noriyuki Matsunaga}
\affiliation{Department of Astronomy, Graduate School of Science, The University of Tokyo, 7-3-1 Hongo, Bunkyo-ku, Tokyo 113-0033, Japan}
\affiliation{Laboratory of Infrared High-resolution spectroscopy (LiH), Koyama Astronomical Observatory, Kyoto Sangyo University, Motoyama, Kamigamo, Kita-ku, Kyoto 603-8555, Japan}
\email{matsunaga@astron.s.u-tokyo.ac.jp}

\author{Shogo Otsubo}
\affiliation{Laboratory of Infrared High-resolution spectroscopy (LiH), Koyama Astronomical Observatory, Kyoto Sangyo University, Motoyama, Kamigamo, Kita-ku, Kyoto 603-8555, Japan}
\email{otsubo@cc.kyoto-su.ac.jp}

\author{Tomomi Takeuchi}
\affiliation{Laboratory of Infrared High-resolution spectroscopy (LiH), Koyama Astronomical Observatory, Kyoto Sangyo University, Motoyama, Kamigamo, Kita-ku, Kyoto 603-8555, Japan}
\email{takeuchi.t@cc.kyoto-su.ac.jp}

\begin{abstract}
The lower level of the \heilabel\ transition is metastable and therefore
well populated. In \hii\ regions, the line can become optically thick,
and its photons then undergo resonant scattering. \citet{Draine2026ApJ} predicted
that scattering splits the profile into several peaks. We present Magellan/WINERED spectra
($R \approx 70{,}000$) at six pointings along a line from the ionizing
source CEN~1 across the southwestern bar of M17, with \heilabel\ and
\pag\ in the same exposures. Every \heilabel\ profile is multi-peaked,
while the \pag\ profiles are single-peaked. Because \pag\ does not undergo resonant scattering, 
it traces gas kinematics; its single peak argues against a purely kinematic origin of the \hei\ peaks, 
and the shared exposures exclude an instrumental origin. 
We therefore attribute the \hei\ peaks to resonant scattering, as predicted. We compare the profiles 
quantitatively with the static, spherically symmetric models of \citet{Draine2026ApJ}, 
using width and asymmetry measures independent of the systemic velocity. 
At two pointings far from CEN~1, no static model of any optical depth reproduces width and asymmetry together: 
both profiles are more red-weighted than static models of comparable width. 
Such profiles cannot arise from uniform motion but are expected from scattering in expanding gas. 
We argue that the profile is shaped by both the kinematics and the optical depth of the He$^{+}$ zone, 
and resolved \heilabel\ profiles could therefore provide a new diagnostic of these properties in \hii\ regions.
\end{abstract}



\section{Introduction} \label{sec:intro}
\setcounter{footnote}{0}

The \heilabel\ emission line, one of the strongest in the near-infrared, is
observed in \hii\ regions, planetary nebulae, and active galactic nuclei. The
lower level of the transition producing this line, i.e., He$^0$ $1s2s$ \term, is
a metastable state. Consequently, despite the excitation potential being as high
as $\sim$20~eV, helium atoms in such an excited state are expected to be highly
abundant in nebular environments, and the \heilabel\ line can become optically
thick, so that its photons are resonantly scattered.

\begin{figure*}[t]
\centering
\gridline{\fig{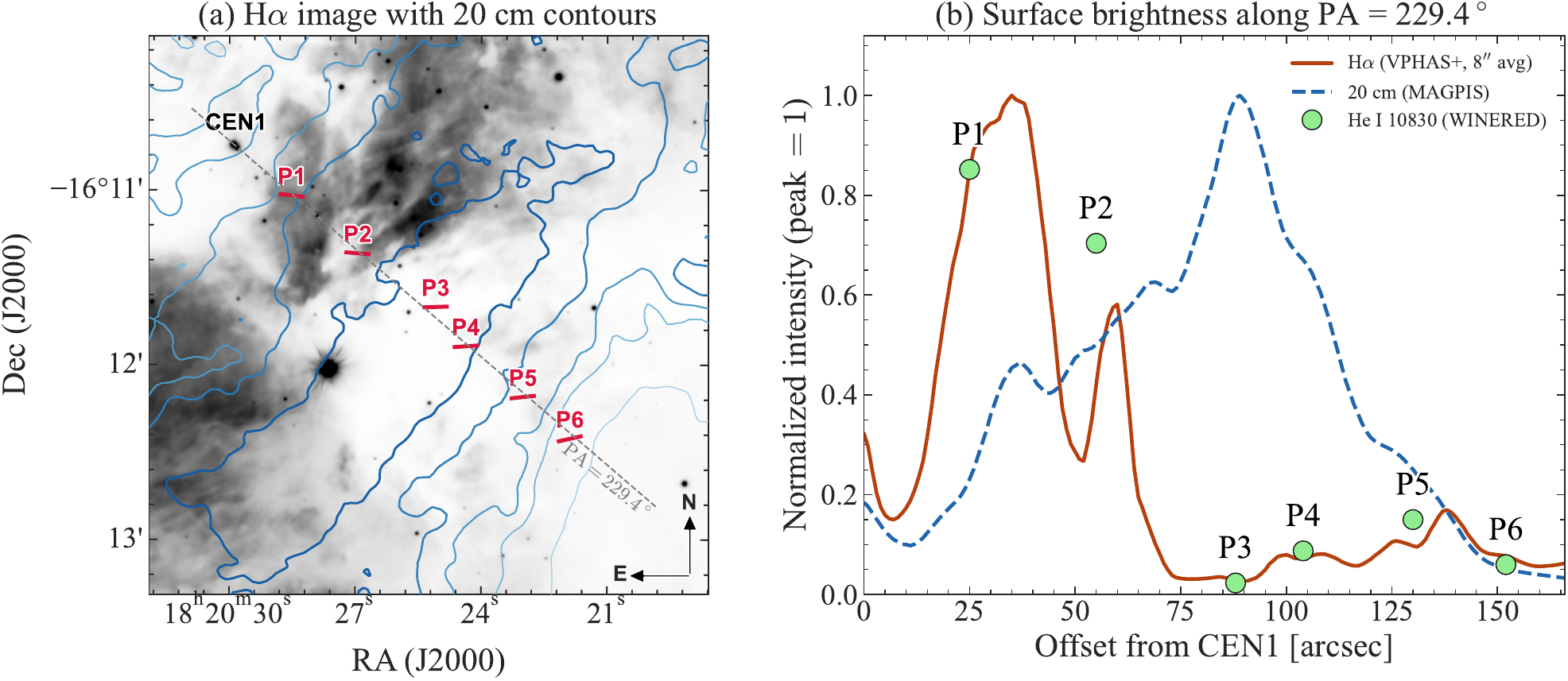}{0.9\textwidth}{}}
\caption{
(a) The six WINERED slit positions, placed along a line at PA $= 229.4^\circ$
from CEN~1 and drawn at the position angle of each exposure
(Table~\ref{tab:obslog}); the slit lengths are to scale, and the widths are
exaggerated for visibility. The background is the VPHAS+ H$\alpha$ image and the
contours are MAGPIS 20~cm (15\,mJy\,beam$^{-1} \times 2^n$). (b) Surface
brightness along the same line: H$\alpha$ and 20~cm as continuous profiles from
the images, and the \hei\ line strengths as points, one per pointing, measured
over $\pm150$~\kms\ about \vlsrpag\ after the pedestal subtraction of
\S\ref{sec:meth:metrics} and, at P3 to P6, interpolation over the OH airglow
interval, and scaled by a single factor fixed at P1. The range of the line
strength among the individual exposures is smaller than the symbols.
}
\label{fig:slits}
\end{figure*}

Using Monte Carlo radiative transfer calculations for a static, spherically
symmetric gas, \citet[hereafter D26]{Draine2026ApJ} predicted a characteristic
distortion of the \heilabel\ profile by resonant scattering. Unlike hydrogen
Ly$\alpha$, which is also strongly affected by resonant scattering, this
behavior arises because the \heilabel\ triplet components with $J=0$, 1, and 2
are located at very closely spaced wavelengths. As a result, as the optical
depth increases, multiple peaks are expected to appear in the \hei\ profile,
accompanied by an overall blueshift. This offers the potential for a new method
of diagnosing the physical conditions in nebulae from emission-line profiles,
but verifying it requires a near-infrared spectrograph with $R \equiv
\lambda/\Delta \lambda \gtrsim 10{,}000$. Such resolution had been reached long
ago: with a Fabry--P\'erot interferometer at $R \approx 36{,}000$,
\citet{Vaughan1968ApJ} found P Cygni-like profiles of the line in planetary
nebulae and the Orion Nebula and attributed them to scattering in expanding gas,
before such profiles had been calculated. Prompted by
\citetalias{Draine2026ApJ}, \citet{Izotov2026AA} found \heilabel\ broader than
H$\beta$ at $R = 5600$ in compact star-forming galaxies and attributed the
broadening in part to resonant scattering and in part to turbulence in denser
inner regions.

M17, also known as the Omega Nebula, is a giant \hii\ region in our Galaxy and
is an ideal target for this test. The dominant ionizing source is CEN~1, a pair
of O4\,V stars \citep{Chini1980A&A,Hoffmeister2008ApJ}, whose hard radiation
field ionizes helium as well as hydrogen. Combined with the high densities of
the southwestern (SW) bar ($n_e \sim 10^2$--$10^3~{\rm cm}^{-3}$;
\citealt{Matthews1979A&A}), this implies a large He$^0$ \term\ population, and
\citetalias{Draine2026ApJ} predicts a large optical depth there, placing M17 in
the regime where profile distortion should be unambiguous. The SW bar is an
ionization front viewed nearly edge-on \citep[hereafter P07]{Pellegrini2007ApJ},
so sight lines through it cross long path lengths of ionized gas. Archival radio
continuum imaging of the region is available (MAGPIS; \citealt{Helfand2006AJ}),
and comparison with H$\alpha$ imaging (VPHAS+; \citealt{Drew2014MNRAS}), which
unlike the radio is attenuated by dust, shows where extinction is strong.

In this paper, we report scattering-induced distortion of the \heilabel\ profile
in M17 and test whether static models of the kind computed by
\citetalias{Draine2026ApJ} can account for it. Section~\ref{sec:obs} describes
the observations and the data reduction. Section~\ref{sec:obsprof} presents the
observed profiles and the evidence that their structure is produced by resonant
scattering. Section~\ref{sec:static} sets out the static models and the
shift-invariant metrics used to compare them with the data, and shows that they
fail at two of the pointings farthest from CEN~1. Sections~\ref{sec:disc} and
\ref{sec:sum} discuss the result and summarize. Wavelengths are given in vacuum,
although line labels such as \heilabel\ follow the common practice of air
wavelengths.


\begin{deluxetable*}{lcccccccc}
\tablecaption{WINERED observation log of the M17 pointings\label{tab:obslog}}
\tablewidth{0pt}
\tablehead{
\colhead{ID} &
\colhead{$\alpha$ (J2000)} &
\colhead{$\delta$ (J2000)} &
\colhead{$d_{\rm CEN1}$\tablenotemark{a}} &
\colhead{UT (start)\tablenotemark{b}} &
\colhead{$t_{\rm exp}$} &
\colhead{Slit PA} &
\colhead{S/N$_{\rm peak}$\tablenotemark{c}} &
\colhead{$v_{\rm LSR}$(\pag)\tablenotemark{d}} \\
\colhead{} & \colhead{} & \colhead{} &
\colhead{[arcsec]} & \colhead{[hh:mm]} & \colhead{[s]} &
\colhead{[deg]} & \colhead{} & \colhead{[\kms]}
}
\startdata
P1 & 18:20:28.52 & $-16$:11:01.9 & 25  & 03:08 & $60 \times 3$ & 83  & 221 & 29.2 \\
P2 & 18:20:26.95 & $-16$:11:21.7 & 55  & 03:02 & $60 \times 3$ & 86  & 315 & 26.0 \\
P3 & 18:20:25.09 & $-16$:11:40.1 & 88  & 02:45 & $60 \times 3$ & 92  & 12  & 15.9 \\
P4 & 18:20:24.37 & $-16$:11:53.6 & 104 & 02:38 & $60 \times 3$ & 94  & 56  & 9.4 \\
P5 & 18:20:23.01 & $-16$:12:11.0 & 130 & 02:31 & $60 \times 3$ & 96  & 87  & 14.1 \\
P6 & 18:20:21.89 & $-16$:12:25.3 & 152 & 02:06 & $60 \times 6$ & 103 & 23  & 25.1 \\
\enddata
\tablenotetext{a}{Projected distance from CEN~1 ($18{:}20{:}29.85$,
$-16{:}10{:}45.0$, J2000) along PA $=229.4\arcdeg$.}
\tablenotetext{b}{All observations were obtained on 2026 June 26 UT.}
\tablenotetext{c}{Peak of the \heilabel\ line divided by the noise at
the peak, per 4~\kms\ bin. The noise, including the photon noise of the
line, is measured from the scatter among the three dither exposures after
removing differences in their overall level (\S\ref{sec:meth:metrics}). At P3 to P6, the OH
airglow interval is interpolated over before the peak is taken
(\S\ref{sec:meth:metrics}).}
\tablenotetext{d}{LSR velocity of the \pag\ profile bisector at the 50\%
level, which serves as the local velocity reference for each pointing
(\S\ref{sec:obs}).}
\end{deluxetable*}

\section{Observations} \label{sec:obs}

The observations were conducted on 2026 June 26 UT, using the WINERED echelle
spectrograph \citep{Ikeda2022PASP} mounted on the 6.5~m Magellan Clay Telescope
at Las Campanas Observatory. WINERED was operated in the HIRES-Y mode with the
$0\farcs3 \times 9\arcsec$ slit, providing a spectral resolving power of $R
\approx 70{,}000$ and covering a wavelength range of 0.96--1.11~\micron.
Observations were carried out at six pointings, P1--P6, placed along a line
extending from the ionizing source CEN~1 at a position angle (PA) of
$229.4^\circ$ (Figure~\ref{fig:slits}a); coordinates and exposures are
summarized in Table~\ref{tab:obslog}. WINERED has no image derotator, so the
slit PA differs between pointings (Table~\ref{tab:obslog}). The pointings were
observed in order from P6 to P1. Sky frames were obtained about 2\arcdeg\ from
M17 in Galactic latitude, three at a time, before and after P6, after P3, and
after P1. The seeing measured by the DIMM was 0\farcs6 to 0\farcs8. Each target
was observed at three dither positions along the slit, offset by $+1\arcsec$,
$-1\arcsec$, and $0\arcsec$.

The data were reduced with the WARP pipeline v1.1 \citep{Hamano2024PASP}, which
performs flat-fielding with exposures of a lamp-illuminated screen,
rectification of the echelle orders, and wavelength calibration. The sky was
subtracted from the raw frames beforehand, because the nebula fills the slit and
the OH airglow doublet at $\lambda \lambda 10834.241,10834.338$\,\AA\
\citep{Oliva2015A&A} overlaps the \heilabel\ profile. We scaled the six raw sky
frames taken before and after each target so that the uncontaminated OH doublet
at $\lambda \lambda 10844.514, 10844.871$\,\AA\ matched its intensity in the
target frame, and subtracted them before the reduction. Because the scaled sky
frames also contain components that do not vary in proportion to the airglow,
mainly persistence from a star observed before the first sky frames, this leaves
a residual in the continuum, of up to 8\% of the line peak (at P6), which we
treat as described in \S\ref{sec:meth:metrics}.

The rectified, wavelength-calibrated two-dimensional spectra of each pointing
were shifted along the slit by the dither offsets and median-combined, and
one-dimensional spectra were extracted over the central 40 pixels
($\sim$6\arcsec) of the slit, rejecting pixels that depart from the spatial
median by more than $4\sigma$ and rescaling the sum to the full aperture. For P1
alone, where the profile varies significantly along the slit, we additionally
extracted spectra in eight bins of 5 pixels each ($\sim$0\farcs75, comparable to
the seeing).

Dividing by the flat field removes most of the echelle blaze function. To
measure what remains, an F-type star was observed immediately after the science
exposures. It shows no significant telluric absorption at the \heilabel\ and
\pag\ wavelengths, so no telluric correction was applied. The residual
sensitivity, derived with the star's intrinsic \pag\ absorption masked and
interpolated over, departs from a constant by less than 0.1\% across a line
profile and by 1.3\% between the two lines, which is immaterial to the profile
shapes. The spectra are presented in counts, without absolute calibration.

\begin{figure*}[t]
\centering
\gridline{\fig{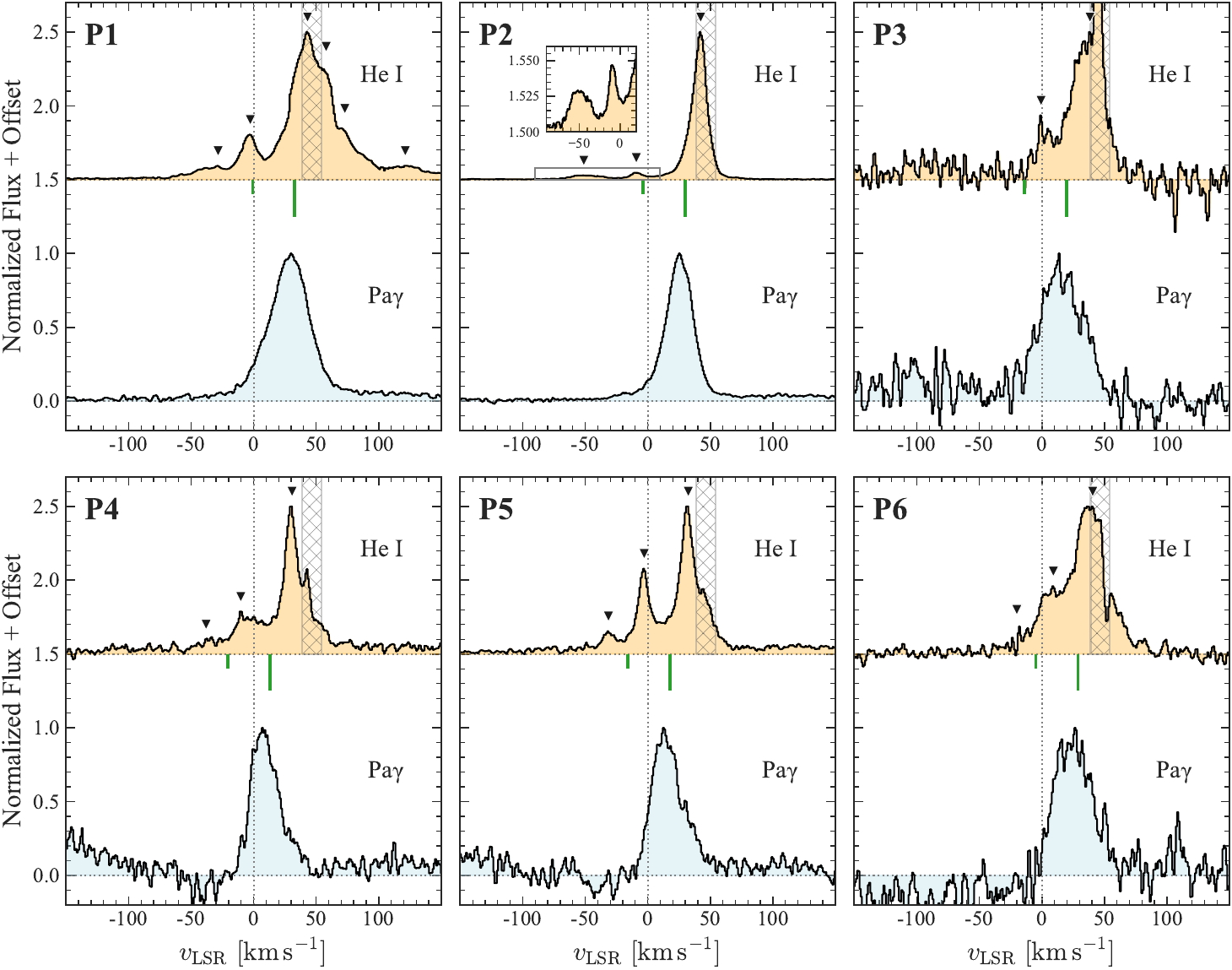}{0.9\textwidth}{}}
\caption{\heilabel\ and \pag\ profiles at the six pointings, ordered by
projected distance from CEN~1. In each panel, the upper trace is \hei\ and the
lower is \pag, each normalized to its own peak and offset vertically for
clarity; velocities are in the LSR frame (\S\ref{sec:obs}). Triangles above the
\hei\ profiles mark the peaks identified by eye. In each panel, the gray
cross-hatched region over the \hei\ trace marks the interval $39 < v_{\rm LSR} <
54$~\kms, which is affected by residuals of the OH airglow after sky subtraction
(\S\ref{sec:obs}). Short green ticks below each \hei\ profile mark the $J=0$ and
$J=1,2$ components of the optically thin multiplet at \vlsrpag, as a guide; the
systemic velocity of the He$^{+}$ zone may differ by a few \kms.
}
\label{fig:profiles}
\end{figure*}

Velocities are referred to a rest wavelength of 10941.082~\AA\ for \pag\ and,
for \heilabel, to the $gf$-weighted multiplet wavelength of 10833.138~\AA,
following \citetalias{Draine2026ApJ}. They are given in the local standard of
rest (LSR) frame throughout, obtained from the wavelength solution with
barycentric and solar-motion corrections applied, the latter adopting the
standard solar motion of 20~\kms\ toward $(\alpha, \delta)=(18^{\rm
h},+30\degr)$ in B1900. Within each pointing, \pag\ also serves as a local
velocity reference. Because its profile is asymmetric, we characterize its
position by the profile bisector, the midpoint between the blue and red sides at
the 50\% level; its LSR velocity, \vlsrpag, is listed in Table~\ref{tab:obslog}.
Velocity intervals that are tied to the line in the measurements below are
specified relative to this bisector. The profile metrics of
\S\ref{sec:meth:metrics} are invariant under a rigid shift and depend only
weakly on the reference, through the placement of the $\pm150$~\kms\ window.


\begin{figure}[t]
\centering
\gridline{\fig{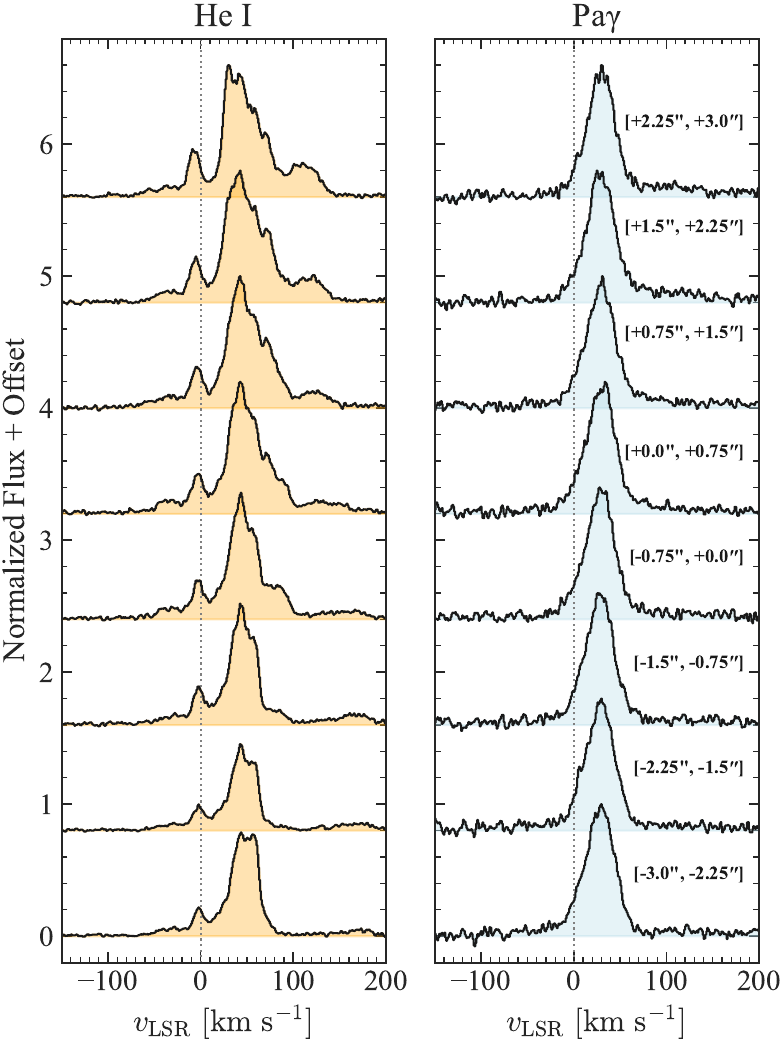}{\columnwidth}{}}
\caption{Variation of the line profiles within the P1 slit, which lies at PA $=
83^\circ$. Spectra extracted from eight independent spatial bins, each 5 pixels
($0\farcs75$) wide and comparable to the seeing, are shown for \heilabel\ (left)
and \pag\ (right), offset vertically in order of position along the slit. Each
profile is normalized to its own peak. Velocities are in the LSR frame; all
eight bins share a single velocity axis.
}
\label{fig:waterfall}
\end{figure}

\section{The Observed \texorpdfstring{\ion{He}{1}}{He I} Profiles} \label{sec:obsprof}

Figure~\ref{fig:profiles} shows the \hei\ and \pag\ emission-line profiles
obtained at the six pointings. Every \hei\ profile is multi-peaked: at least two
peaks are resolved at each position, three at P2, P4, P5, and P6, and more at
P1. At P5, troughs lie near the wavelengths of both the $J=0$ and $J=1,2$
components at \vlsrpag\ (Figure~\ref{fig:profiles}), indicating that each
component is split, as \citetalias{Draine2026ApJ} predicted for high \term\
column densities. The \pag\ profiles, observed simultaneously through the same
slit and reduced identically, remain single-peaked at every position. The lower
level of \pag\ ($n=3$) is so sparsely populated that the optical depth of the
line is negligible,\footnote{For Case~B recombination at $T_e = 10^{4}$~K,
$\tau(\mathrm{Pa}\gamma) \approx 8\times10^{-10}\,(n_e/10^{3}~{\rm
cm^{-3}})^{2}\,(L/1~{\rm pc})\,(b/10~\mathrm{km~s^{-1}})^{-1}$ at line center
over a path $L$.} whereas that of \heilabel\ in M17 is of order
10$^{2}$--10$^{3}$ \citepalias{Draine2026ApJ}; \pag\ therefore does not undergo
resonant scattering and traces the velocity field of the gas almost directly.
Moreover, hydrogen is ionized wherever helium is, so \pag\ samples all of the
gas that emits \hei. The absence of corresponding peaks in \pag\ therefore
argues against a purely kinematic origin for the \hei\ peaks. The shared
exposures and reduction likewise rule out an instrumental or calibration origin.

Along the slit, the distortion changes little at most pointings but markedly at
P1. Across eight independent bins there, each comparable to the seeing, the
shape of the \hei\ profile changes markedly, while the \pag\ profile extracted
from the same bins is nearly unchanged (Figure~\ref{fig:waterfall}). Because
\pag\ is emitted along the same sight lines in each bin and is nearly unchanged,
this variation cannot be explained by kinematics alone, and it is further
evidence for resonant scattering. P1, however, lies on the steep flank of a
compact H$\alpha$ ridge (Figure~\ref{fig:slits}a), and its sight lines evidently
cross a complex gas structure; we leave its physical interpretation to future
work.

To our knowledge, earlier observations of the line
\citep{Vaughan1968ApJ,Izotov2026AA} did not resolve the splitting of each
multiplet component; the present observations resolve the splitting at P5 and,
with an optically thin comparison line observed in the same exposures, show that
it arises from resonant scattering.

Beyond these common features, the profiles differ from sight line to sight line,
in shape and in strength. At P2, the profile is dominated by a single bright
peak with an asymmetric blue wing and two weak peaks blueward of it; \pag\ shows
a smooth tail extending to $v_{\rm LSR} \approx -30$~\kms; at this pointing, the
scattered and kinematic contributions cannot be separated with the present data.
In strength, the \hei\ line at P3 and P4 is weak relative to the radio
continuum, as is H$\alpha$ (Figure~\ref{fig:slits}b), consistent with the heavy
obscuration of the SW bar \citepalias{Pellegrini2007ApJ}. Obscuration lowers the
signal-to-noise ratio, which is lowest at P3 (Table~\ref{tab:obslog}), but being
gray across a line profile, it can neither generate nor remove the structure
discussed above.

The \pag\ velocities themselves differ between sight lines. \vlsrpag\ ranges
from $+9.4$ to $+29.2$~\kms\ (Table~\ref{tab:obslog}), all redshifted,
consistent with the recession of M17 as a whole \citepalias{Pellegrini2007ApJ}.
The pattern is not monotonic with projected distance from CEN~1; together with
the asymmetry of the \pag\ profiles, it shows that more than one kinematic
component lies along at least some sight lines.


\section{Comparison with Static Models} \label{sec:static}

The splitting of the profile confirms the prediction of
\citetalias{Draine2026ApJ} qualitatively. That prediction was made for a static,
uniform sphere, and we now ask whether such models reproduce the observed
profile shapes quantitatively.

\subsection{The Model Grid} \label{sec:meth:static}

\citetalias{Draine2026ApJ} treats resonant scattering in a static, spherically
symmetric, uniform medium by Monte Carlo radiative transfer, and we use the
emergent spectra made publicly available with that work
\citep{Draine2026Dataverse}. Following \citetalias{Draine2026ApJ}, throughout
this paper, $\tau_{\rm tot}$ denotes the sum of the line-center optical depths
of the $J=0$, 1, and 2 components, and $b$ is the Doppler parameter of the
scattering atoms. We use the dust-free models with $b = 6.45$, 10, 15, and
20~\kms, which cover, at each $b$, 33 values of $\tau_{\rm tot}$ from $0.01$ to
$10^{6}$, so the grid extends from the optically thin limit to optical depths
far beyond the $\tau_{\rm tot} = 2000$ that \citetalias{Draine2026ApJ} estimates
for the radio continuum peak of the SW bar (M17-B in
\citetalias{Draine2026ApJ}), which corresponds to P3. Before comparison with the
data, the model spectra are convolved with a Gaussian matched to the spectral
resolution, $R=70{,}000$. We refer to the tabulated models as the grid and to
the continuous set obtained by interpolating between them as the static family.

\subsection{Shift-invariant Profile Metrics} \label{sec:meth:metrics}

Comparing the static grid with the observations requires a measure of profile
shape, and the obvious choice is unavailable here: \citetalias{Draine2026ApJ}
proposed the mean blueshift as an observational diagnostic, but the asymmetry of
the observed \pag\ profile makes the systemic velocity difficult to fix, and any
measure referred to it inherits that uncertainty. We therefore characterize each
profile by its cumulative flux distribution, which can be described without a
velocity zero point. Writing $v_q$ for the velocity at which the cumulative flux
reaches $q\%$ of the total, we use two widths,
\begin{equation}
  W_{80} = v_{90} - v_{10}, \qquad W_{50} = v_{75} - v_{25},
\end{equation}
the first sensitive to the wings of the profile and the second to its core,
together with the asymmetry parameter of \citet{Liu2013MNRAS},
\begin{equation}
  A = \frac{(v_{90} - v_{50}) - (v_{50} - v_{10})}{W_{80}},
\end{equation}
which is negative when the median lies redward of the midpoint of the
$v_{10}$--$v_{90}$ interval, that is, when the flux is concentrated to the red
with a tail to the blue (a red-weighted profile); it vanishes for a symmetric
profile and is bounded by $\pm1$. All three quantities are invariant under a
rigid velocity shift.

Identical estimators are applied to the models and to the data: the spectrum is
rebinned to 4~\kms, negative values, which arise from noise, are set to zero so
that the cumulative flux rises monotonically, and the flux is accumulated over
$\pm150$~\kms\ about \vlsrpag\ for the data and about the multiplet centroid for
the models. A constant pedestal is subtracted from each observed spectrum,
estimated from the line wings 100--150~\kms\ on either side of \vlsrpag; at P1,
the blue wing alone is used, the red wing of \hei\ there containing a component
with no counterpart in \pag\ or at the other pointings. The interval $39 <
v_{\rm LSR} < 54$~\kms, affected by residuals of the OH airglow, is not
interpolated over, because doing so in the models would replace a narrow peak at
its edge by a linear ramp that mimics a red wing. An OH residual of $\pm10\%$ of
the doublet changes $D$ by at most 15\% at P4 and P5. At P3, the residual
exceeds the line itself, and the comparison is not made. Because setting
negative values to zero biases the metrics of faint lines, the models are
compared as they would be observed: each model spectrum, scaled to the observed
line flux, is given the noise of the observation and processed identically, and
$\mathbf{m}_{\rm s}$ below is the mean of 200 such realizations. Because the
systemic velocity is uncertain, the models are also shifted by up to
$\pm8$~\kms, and the smallest $D$ is adopted.

Uncertainties are obtained from 1000 realizations of each spectrum with Gaussian
noise added. The noise variance in a pixel of flux $f$ is modeled as $a + b f$,
with $a$ and $b$ fitted to the scatter among the three dither exposures; the
noise is then scaled by 1.5 to account for the correlation between adjacent
pixels. The three metrics differ in units and in precision, and because they are
built from the same quantiles, their errors are correlated; the realizations
therefore give the full covariance matrix $\mathbf{C}$ of $(W_{80}, W_{50}, A)$.
The separation of an observed point $\mathbf{m}$ from the static family is
measured by the Mahalanobis distance,
\begin{equation}
  D = \min_{b,\,\tau_{\rm tot}}
  \left[(\mathbf{m}-\mathbf{m}_{\rm s})^{\rm T}\,\mathbf{C}^{-1}\,
  (\mathbf{m}-\mathbf{m}_{\rm s})\right]^{1/2},
\end{equation}
where $\mathbf{m}_{\rm s}(b,\tau_{\rm tot})$ are the metrics of the static
family; $D$ is the separation from the nearest static model in units of the
measurement error in that direction. For a point drawn from a single model,
$D^{2}$ follows a $\chi^{2}$ distribution with three degrees of freedom, and we
quote the equivalent Gaussian significance on that basis. This is conservative,
because minimizing over a two-parameter family leaves fewer degrees of freedom.
Scaling the noise by 1.65 instead of 1.5 lowers the significances by at most
$1.8\sigma$.

\subsection{Results} \label{sec:res:static}

\begin{figure*}[t]
\centering
\gridline{\fig{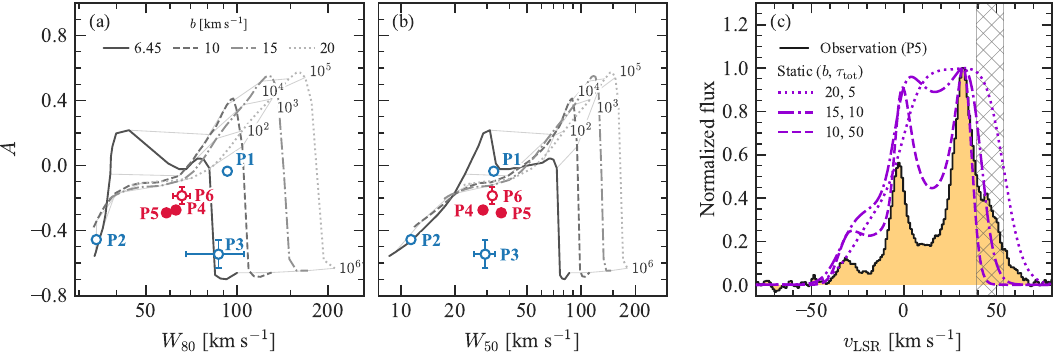}{0.95\textwidth}{}}
\caption{
(a), (b) The six pointings and the static models of \citetalias{Draine2026ApJ}
in the planes ($W_{80}, A$) and ($W_{50}, A$). Black and gray curves,
distinguished by line style, join models of equal $b$ over $\tau_{\rm
tot}=0.01$--$10^{6}$; thin lines join models of equal $\tau_{\rm tot}$ at each
decade from 1 to $10^{6}$, labeled by $\tau_{\rm tot}$. Filled red circles are
P4 and P5, which depart from the static family, and the open red circle is P6;
open blue circles are P1, P2, and P3, which are treated separately (see text).
P3 is plotted with the airglow interval interpolated over. The curves show the
models without noise; the separations quoted in the text are measured from the
models given the noise of each pointing (\S\ref{sec:meth:metrics}). Error bars
are $1\sigma$. (c) The \heilabel\ profile at P5, as extracted, and the three
grid models nearest to it in the three metrics (purple; $b=20$, 15, and 10~\kms\
with $\tau_{\rm tot}=5$, 10, and 50, in the line styles of panels a and b). Each
profile is normalized to its peak, and the models are shifted to align the
reddest peaks. The hatched interval is affected by residuals of the OH airglow
(\S\ref{sec:meth:metrics}).
}
\label{fig:metrics}
\end{figure*}

Static models produce asymmetric profiles of either sign \citepalias[see
Figure~6b of][for $b = 6.45$~\kms\ and $\tau_{\rm tot} \leq
10^{4}$]{Draine2026ApJ}. At low optical depth, the $J=0$ component, 34~\kms\
blueward of the $J=1,2$ pair, makes $A$ negative; as the optical depth rises,
scattering moves weight into $J=0$, raising $A$, while both widths grow. For
$W_{80} = 55$--70~\kms, the static family thus has $A \gtrsim -0.13$
(Figure~\ref{fig:metrics}); only at $\tau_{\rm tot} \gtrsim 2\times10^{5}$ does
$A$ fall again, to about $-0.7$, with $W_{50} > 75$~\kms.

P4 and P5 lie outside this family, on its red-weighted side. They have $A =
-0.27 \pm 0.03$ and $-0.29 \pm 0.01$ at $W_{80} = 63$ and 58~\kms, far more
red-weighted than any static model of similar width; their separations are $D =
12$ and 15 ($12\sigma$ and $15\sigma$). P6, one of the two faintest, lies within
the family ($D = 1.0$), although, as at P5, its reddest peak is by far the
strongest, unlike in the nearest models; the metrics are insensitive to this
difference. P3, the faintest, is not compared, because residuals of the OH
airglow exceed the line in the airglow interval (\S\ref{sec:meth:metrics}).

Figure~\ref{fig:metrics}c shows what the separation means for the profile at P5.
The three nearest models of the grid lie at similar separations ($D = 17$--18)
along the degeneracy between $b$ and $\tau_{\rm tot}$. In each, however, the two
strongest peaks are of nearly equal height or merge into a flat top, whereas in
the observed profile, the reddest peak is clearly the stronger.

P1 and P2 are treated separately. At P1, the profile changes within the slit
(\S\ref{sec:obsprof}), so the integrated spectrum combines sight lines of
different character and does not test a single static medium; no static model
with a core as broad as that at P1 ($W_{50} = 33$~\kms) has wings as broad
($W_{80} = 93$~\kms, against at most 64~\kms). P2 lies near the static family,
close to the models with $b = 6.45$~\kms\ and $\tau_{\rm tot}$ of a few ($D =
2.4$). Its main peak falls in the airglow interval, and an OH residual of
$\pm10\%$ of the doublet would change $D$ to between 0.7 and 4.5. Its profile,
however, is not reproduced by those models. They have a distinct peak 34~\kms\
blueward of the main peak, from the $J=0$ component, and none beyond it. In the
observed profile, the $J=0$ component is not resolved from an asymmetric blue
wing of the main peak, and two weak peaks lie 50 and 90~\kms\ blueward of it, at
5\% and 2.5\% of its height. The metrics, being integrals over the profile, are
insensitive to structure this weak, so their agreement with models of low
optical depth does not establish that the line at P2 is unaffected by resonant
scattering.


\section{Discussion} \label{sec:disc}

The evidence presented here rests on profile shape rather than on line
intensity, so it requires no absolute flux calibration and is not degenerate
with extinction or with the extent of the He$^{+}$ zone. The shape, however, is
not that of a static, uniform sphere of any optical depth (\S\ref{sec:static}),
so at least one assumption of those models fails in M17.
\citetalias{Draine2026ApJ} noted that such deviations would not be surprising,
because the stars and gas of real \ion{H}{2} regions may be distributed in a
complex way and gas flows may produce velocity gradients.

Of these possibilities, the data favor a velocity gradient within the scattering
gas. A uniform motion cannot explain the departure, because it would shift the
profile without changing its shape or the metrics. Both P4 and P5 have more
negative $A$ than static models of comparable width, i.e., their flux is more
strongly weighted to the red. Such red-weighted profiles are expected when
resonance-line photons scatter in expanding gas, as \citet{Vaughan1968ApJ}
argued for this line and as found for Ly$\alpha$ in outflows
\citep{Verhamme2006A&A}, and our preliminary radiative transfer calculations for
an expanding shell confirm that $A$ then becomes more negative. Finally, the
ionized layer of M17 is independently inferred to expand relative to the neutral
shell and the molecular cloud \citepalias{Pellegrini2007ApJ}. We present the
calculations, with a fit to the profile at P5 and the optical depth it implies,
in a second paper.

Other departures from the models are not excluded. The comparison treats each
profile as the emission of a single scattering medium, whereas more than one
kinematic component lies along at least some sight lines (\S\ref{sec:obsprof});
a superposition of static components at different velocities could alter the
metrics, although it would have to produce in \hei\ the peaks that are absent
from \pag. The models are spherical and uniform, and their spectra are
integrated over the whole sphere, whereas each observation samples a single line
of sight.

The optical depth of the \term\ level matters beyond the \heilabel\ line itself.
Determinations of the primordial helium abundance ($Y_{\rm p}$) from \hei\
recombination lines correct the line intensities for this optical depth, which
is not observed but fitted together with the density and temperature
\citep{Izotov2014MNRAS,Aver2015JCAP,Aver2026ApJ}, using radiative transfer
corrections computed for a spherical nebula without systematic velocity
gradients \citep{Benjamin2002ApJ,Kurichin2026PhRvD}. The \heilabel\ profile is
shaped by this optical depth directly, and at two of the six pointings, models
without velocity gradients fail to reproduce it. Resolved profiles of this line
therefore offer an independent check on those corrections.

\section{Summary} \label{sec:sum}

\begin{enumerate}
    \item We resolve the \heilabel\ profile in the \hii\ region M17 into the
        multiple peaks expected from resonant scattering. At six pointings in
        M17 (P1 to P6, in order of distance from CEN~1), the \hei\ profiles are
        multi-peaked while \pag, observed in the same exposures, is
        single-peaked.

    \item Static scattering models do not account for the profiles at P4 and P5,
        two of the four pointings farthest from CEN~1. In the shift-invariant
        metrics $(W_{80}, W_{50}, A)$, these lie outside the static family of
        \citetalias{Draine2026ApJ} by $12\sigma$ and $15\sigma$. At P6, the
        metrics do not distinguish the profile from static models, and at P3,
        residuals of the OH airglow preclude the comparison.

    \item We argue that the discrepancy reflects a velocity gradient within the
        scattering gas, because both P4 and P5 are more red-weighted than static
        models of comparable width, as expected for scattering in expanding gas.
        The ionized layer of M17 is independently inferred to expand. The
        \heilabel\ profile is thus shaped by both the kinematics and the optical
        depth of the He$^{+}$ zone, and resolved profiles could provide a new
        diagnostic of these properties in \hii\ regions.
\end{enumerate}

\begin{acknowledgments}

We thank B. T. Draine for making the model spectra of D26 publicly available. We
also thank the staff of Las Campanas Observatory for their support during the
observations, and the staff of Carnegie Observatories, especially Andy McWilliam
and Shreyas Vissapragada, for enabling the operation of WINERED.

This paper includes data gathered with the 6.5 meter Magellan Telescopes located
at Las Campanas Observatory, Chile. WINERED was developed by the University of
Tokyo and the Laboratory of Infrared High-resolution spectroscopy (LiH), Kyoto
Sangyo University, with financial support from JSPS KAKENHI (grant
Nos.~16684001, 20340042, 21840052, and 26287028), and from the MEXT Supported
Program for the Strategic Research Foundation at Private Universities
(Nos.~S0801061 and S1411028).

The National Radio Astronomy Observatory is a facility of the National Science
Foundation operated under cooperative agreement by Associated Universities, Inc.
Based on data products from observations made with ESO Telescopes at the La
Silla Paranal Observatory under programme ID 177.D-3023, as part of the VST
Photometric H$\alpha$ Survey of the Southern Galactic Plane and Bulge (VPHAS+,
www.vphas.eu). This research has made use of NASA's Astrophysics Data System.

H.S. is supported by JSPS KAKENHI Grant Numbers JP22K03683, JP19K03917. The 2026
June observing run was partly supported by the TMU Research Fund for Young
Scientists from Tokyo Metropolitan University.

During the preparation of this work, the authors used Claude (Anthropic) in
planning the observations, in writing analysis and plotting code, and in
drafting and editing the text. The authors verified the code and its results,
reviewed and edited the text, and take full responsibility for the content of
the publication.

\end{acknowledgments}

\contribution{
H.S. planned the observations and supervised them remotely, carried out the data
analysis, and wrote the manuscript. D.T. and N.M. contributed to the observing
proposal and the interpretation of the results. D.T., S.O., and T.T. operated
the instrument at the telescope. All authors discussed the results and commented
on the manuscript.
}
\facilities{Magellan:Clay (WINERED), VLA, Effelsberg, ESO:VST (OmegaCAM).}
\software{WARP v1.1, astropy \citepalias{astropy2022}, numpy \citep{Harris2020Natur},
scipy \citep{Virtanen2020NatMe}, matplotlib \citep{Hunter2007CSE}, seaborn
\citep{Waskom2021JOSS}, Claude (Anthropic).}

\bibliography{m17}
\bibliographystyle{aasjournalv7}

\end{document}